\documentclass[preprint]{aastex61}
\usepackage{amsmath}
\usepackage{algorithm, algorithmicx} 
\usepackage[noend]{algpseudocode}

\shorttitle{GPU-Based High Performance Imaging for Mingantu Spectral RadioHeliograph}
\shortauthors{Mei et al.}

\begin{document}

\title{GPU-Based High-Performance Imaging for Mingantu Spectral RadioHeliograph}

\correspondingauthor{Feng Wang}
\email{wangfeng@acm.org, cnwangfeng@gmail.com}

\author{Ying Mei} 
\affiliation{Yunnan Observatories, Chinese Academy of Sciences\\
P.O.Box 110, Kunming 650011, China}
\affiliation{Center of Astrophysics, Guangzhou University\\ No.230, Waihuanxi Rd, Panyu District, Guangzhou 510006, China}
\affiliation{Key Lab of Computer Technology Application of Yunnan Province, Kunming University of Science and Technology\\
No.727 South Jingming Rd., Chenggong District, Kunming 650500, China}
\affiliation{University of Chinese Academy of Sciences\\
19A Yuquan Rd., Shijingshan District, Beijing 100049, China}

\author{Feng Wang}
\affiliation{Yunnan Observatories, Chinese Academy of Sciences\\
P.O.Box 110, Kunming 650011, China}
\affiliation{Center of Astrophysics, Guangzhou University\\
No.230, Waihuanxi Rd, Panyu District, Guangzhou 510006, China}
\affiliation{Key Lab of Computer Technology Application of Yunnan Province, Kunming University of Science and Technology\\
No.727 South Jingming Rd., Chenggong District, Kunming 650500, China}
\affiliation{University of Chinese Academy of Sciences\\
19A Yuquan Rd., Shijingshan District, Beijing 100049, China}

\author{Wei Wang}
\affiliation{National Astronomical Observatory, Chinese Academy of Sciences\\
20A Datun Rd., Chaoyang District, Beijing 100012, China}

\author{Linjie Chen}
\affiliation{National Astronomical Observatory, Chinese Academy of Sciences\\
20A Datun Rd., Chaoyang District, Beijing 100012, China}

\author{Yingbo Liu}
\affiliation{Yunnan Academy of Scientific $\&$ Technical Information\\
No.246 People's East Rd., Kunming 650051, China}

\author{Hui Deng}
\affiliation{Center of Astrophysics, Guangzhou University\\ No.230, Waihuanxi Rd, Panyu District, Guangzhou 510006, China}
\affiliation{Key Lab of Computer Technology Application of Yunnan Province, Kunming University of Science and Technology\\
No.727 South Jingming Rd., Chenggong District, Kunming 650500, China}

\author{Wei Dai}
\affiliation{Key Lab of Computer Technology Application of Yunnan Province, Kunming University of Science and Technology\\
No.727 South Jingming Rd., Chenggong District, Kunming 650500, China}

\author{Cuiyin Liu}
\affiliation{Key Lab of Computer Technology Application of Yunnan Province, Kunming University of Science and Technology\\
No.727 South Jingming Rd., Chenggong District, Kunming 650500, China}

\author{Yihua Yan}
\affiliation{National Astronomical Observatory, Chinese Academy of Sciences\\
20A Datun Rd., Chaoyang District, Beijing 100012, China}

\begin{abstract}

As a dedicated solar radio interferometer, the MingantU SpEctral RadioHeliograph (MUSER) generates massive observational data in the frequency range of 400 MHz -- 15 GHz. High-performance imaging forms a significantly important aspect of MUSER's massive data processing requirements. 
In this study, we implement a practical high-performance imaging pipeline for MUSER data processing.
At first, the specifications of the MUSER are introduced and its imaging requirements are analyzed. Referring to the most commonly used radio astronomy software such as CASA and MIRIAD, we then implement a high-performance imaging pipeline based on the Graphics Processing Unit (GPU) technology with respect to the current operational status of the MUSER. A series of critical algorithms and their pseudo codes, i.e., detection of the solar disk and sky brightness, automatic centering of the solar disk and estimation of the number of iterations for clean algorithms, are proposed in detail.
The preliminary experimental results indicate that the proposed imaging approach significantly increases the processing performance of MUSER and generates images with high-quality, which can meet the requirements of the MUSER data processing.

\end{abstract} 

\keywords{Astronomical instrumentation: interferometric array -- the Sun: interferometer imaging -- techniques: imaging pipeline -- methods: solar disk and sky brightness -- methods: clean}

\section{Introduction} \label{sec:intro}

Radio astronomical observation techniques have undergone rapid advancements since Martin Ryle first proposed the use of aperture synthesis imaging for radio astronomy \citep{ryle1962new}. Modern interferometric instruments ranging from the Expanded Very Large Array (EVLA), the Low Frequency Array (LOFAR), and the Australian Square Kilometre Array Pathfinder (ASKAP) to the world's largest radio telescope Square Kilometre Array (SKA) are increasingly subject to a data deluge, which poses great challenges to high-performance data processing. 

Over the past few years, observations of solar emissions at metric wavelengths have yielded new insights into solar activities such as particle acceleration, coronal mass ejections (CMEs) and flares. The currently available radio telescopes observes the solar at only at a few discrete frequencies: the Nobeyama Radioheliograph (NoRH\footnote{\url{http://solar.nro.nao.ac.jp/norh/}}) makes observations at 17 GHz and 34 GHz, the Nancay Radioheliograph (NRH) operates at 5 -- 10 frequencies in the range of 150 -- 450 MHz \citep{kerdraon1997nanccay}, the frequency range of Gauribidanur Radioheliograph is 40 -- 150 MHz, and the Siberian Solar Radio Telescope (SSRT\footnote{\url{http://en.iszf.irk.ru/The_Siberian_Solar_Radio_Telescope}}) acquires data at 5.7GHz. Meanwhile, large interferometers such as the Very Large Array \citep{napier1983very} and the Westerbork Synthesis Radio Telescope (WSRT) are primarily designed for rotational synthesis observations and their fields of views are typically too small to cover the entire solar disk \citep{baars1973synthesis, yan2015first}.

The MingantU SpEctral RadioHeliograph (MUSER) is a solar-dedicated interferometric array constructed to investigate the dynamics properties of solar bursts. The MUSER was built at Mingantu in Inner Mongolia, China. 
The MUSER comprises two sub-arrays: MUSER-I, operating in the frequency range of 400 MHz -- 2 GHz with 40 4.5-m antennas, whose observation (covers 1.6 GHz bandwidth, two polarizations) is divided into four bands, and MUSER-II, operating in the frequency range of 2 GHz -- 15 GHz with 60 2-m antennas, whose observation (covers 13 GHz bandwidth, two polarizations) is divided into 33 bands. Notice that there are 8 repeated channels among the 528 channels of MUSER-II. The specifications of the MUSER are listed in Table~\ref{basicparameters} \citep{yan2011progress}. The digital receivers can be specified to work in loop mode (loop both in bands and polarizations) or fix mode (fix in one band and one polarization). Every 3 ms, both the digital receivers of MUSER-I and MUSER-II receive 400 MHz analog signals in a specific polarization and the correlators simultaneously output 16 channels with a 25 MHz bandwidth for each channel \citep{yan2011progress,Wang2015Distributed}. 
Therefore, the MUSER is expected to provide a new observational window on solar activities as a new instrument that is capable of operating at hundreds of frequencies and nearly simultaneously generating high-quality radio images with high temporal, high spatial, and high spectral resolutions. \citep{YAN204,YAN2009,YAN2010,yan2011progress}. 

\begin{deluxetable*}{ccc}[htbp]
\tablecaption{MUSER Characteristic and Performance} 
\label{basicparameters}
\tablecolumns{3}
\tablenum{1}
\tablewidth{0pt}
\tablehead{
\colhead{Parameters} &
\colhead{MUSER-I} &
\colhead{MUSER-II}
}
\startdata
  Antenna array            & 40 (4.5 meters)    &  60 (2 meters)\\
  Frequency range      &  0.4 GHz$-$2.0 GHz &  2.0 GHz$-$15 GHz\\
  Frequency resolution   &  64 channels   &  520 channels\\
  Channel bandwidth    & \multicolumn{2}{c}{25 MHz}\\
  Time resolution      & 25 ms & 206.25 ms\\
  Spatial resolution   & ${51.6}''- {10.3}''$ & ${10.3}''- {1.4}''$\\
  Baseline length      & $\sim8 m - \sim3000 m $  & $ \sim4 m - \sim3000 m$ \\
  Polarization         & \multicolumn{2}{c}{Dual circular left and right}\\
  Dynamic range        & \multicolumn{2}{c}{25 db (snapshot)}\\
  Field of view        & \multicolumn{2}{c}{$0.6^{\circ} - 7^{\circ}$}\\
  \hline
\enddata
\end{deluxetable*}

Since the successful installation of the MUSER hardware, a series of measurements and experiments has been conducted on the MUSER system \citep{wang2013calibration}. The phase closure relationship in the experiments reflected the appropriate design of the whole system \citep{wang2013calibration}. The first radio burst presented by \cite{yan2015first} and the simulation results by \cite{du2015image} demonstrated the imaging capability of the MUSER. Meanwhile, a distributed computing framework, Opencluster, has been designed to meet the requirements of high-performance data processing \citep{Wang2015Distributed, wei2016opencluster}. 

However, high-performance imaging is still a critical problem for the MUSER. 
According to the specific parameters listed in Table~\ref{basicparameters}, both MUSER-I and MUSER-II generate raw observational data frames that include 16 frequencies (one band) with one specified polarization every 3 ms, which implies that 614,400 images need to be generated in 1 min if needed. Although the Graphics Processing Unit (GPU) technology has been used in previous studies \citep{Wang2015Distributed}, the performance of the MUSER further needs to be boosted urgently. 

In addition, due to certain constraints of the current MUSER system, some specific algorithms should be further studied. For example, as a geosynchronous satellite is used as the only available calibrator source and its phase-tracking accuracy is insufficient, it is required to further correct the phase error in dirty images by detecting the solar disk and moving the disk to the image center. 

In this study, we focus on the implementation of the MUSER high-performance imaging procedure, particularly the dedicated processing approaches for the MUSER in its current status. The rest of the paper is organized as follows. In section 2, a review of related work is presented. After presenting a broad overview of the imaging pipeline in section 3, we discuss key steps such as the estimation of the solar disk and sky brightness, solar disk automatic centering and the improved clean algorithm in section 4. The implementation and imaging performance is analyzed in section 5. Section 6 discusses the existing problems, and finally, section 7 concludes the paper along with the presentation of certain aspects of our future work.

\section{Related Work}

A fundamental theory of radio interferometric imaging is the existence a Fourier transform (3D) between the sky brightness and the visibility data obtained with a radio interferometric array \citep{Thompson2008, taylor1999synthesis}. 
In particular, if all the visibility data can be approximately considered to lie on a plane and if the field of view is sufficiently small or the time of observation is sufficiently small (snapshot observations), the computation of the sky brightness can be reduced to a 2D transform \citep{perley1999imaging, bhatnagar1999wide}. 
Although ignoring the w-term can effectively reduces the load during imaging computing, careful calculation of the phase error caused by it is needed in the general case as which may bring serious phase error in images\citep{perley1999imaging}.

In general, gridding, weighting, Fourier transform and dirty image cleaning form the main steps in processing data from the interferometer. Radio antenna arrays typically have non-uniform spacing, which gives them simultaneous access to low and high frequencies, but with sparse sampling in the UV plane. In practice, the data are required to be variously weighted according to their reliability \citep{sault2007imaging}. Further, weighting functions are used for controlling the synthesized beam shape \citep{perley1999imaging, rau2009advances}. Different methods of weighting, such as natural weighting, uniform weighting and robust weighting, are used during imaging. Gridding is the convolution of the visibilities with a particular function, which resamples the visibility data at grid points \citep{Thompson2008} as the fast Fourier transform (FFT) requires the visibilities to lie on a regular rectangular grid. 

There is a series of clean algorithms for removing instrumental artifacts in dirty images directly generated from the gridding and FFT steps. In this context, H{\"o}gbom presented a classic image deconvolution algorithm that was considered a milestone in radio astronomy \citep{hogbom1974aperture}. Subsequently, many improved clean algorithms have been proposed \citep{taylor1999synthesis, Thompson2008}. 
For solar observations, the Steer clean was proposed to solve the problems of failing to restore diffuse features and the excessive computational time of the standard H{\"o}gbom clean. The Steer algorithm suitably restores complicated brightness structures in daily solar images of the Nobeyama
Radioheliograph, consuming only one-tenth the time of the classical H{\"o}gbom algorithm \citep{koshiishi2003restoration}. Although 
H{\"o}gbom algorithm can address point sources and extended sources very well, the Steering clean is typically regarded as a more appropriate algorithm for cleaning the dirty image of the Sun.

The processing performance forms another key issue in the data reduction of interferometer. To accelerate the computational speed of the gridding and clean steps, a GPU is widely used for interferometric imaging. A GPU-based approach for gridding makes the w--projection 100 times faster than that of general CPU-based imaging tools \citep{muscat2014high}. The thread coarsening based on a GPU is applied in MeerKAT \citep{brederode2016meerkat} to improve the efficiency of the existing convolutional gridding algorithm \citep{merry2016faster}.There is also the GPU-accelerated software simulator OSKAR \citep{mort2016analysing}, which makes it possible to run large-scale, full-sky simulations of aperture arrays on reasonable timescales, and this simulator can be useful for data processing of the SKA. Obviously, the application of GPU--CUDA (Compute Unified Device Architecture) framework has greatly improved the data processing efficiency, and its use is gradually expected to become a trend in massive astronomical data processing in the future. 

In addition to fundamental researches on interferometric imaging, many software packages such as CASA (the Common Astronomy Software Applications) and MIRIAD (Multichannel
Image Reconstruction, Image Analysis and Display) have been developed and widely used for interferometric data processing. CASA has been mainly developed by C/C++ and partially supports the message passing interface (MPI) parallel computing environment. The MIRIAD was initially implemented in FORTRAN-77 and miriad-python \citep{williams2012rapid} was developed to provide Python programming language to the MIRIAD data reduction system. Nevertheless, to handle the data with self-defined data format and meet the unique characteristics of the MUSER, a lightweight and high-efficiency data processing pipeline with appropriate algorithms is in pressing need. 

In summary, imaging is a classical problem for which many algorithms have been proposed previously. However, only a few studies have discussed the interferometric imaging of the Sun and the clean algorithms for extended source. Meanwhile, very few studies have described the implementation of high-performance imaging on a GPU platform in detail, and performance-involved factors such as the detection of solar disk and sky brightness and the determination of the number of iterations for clean algorithms are seldom discussed. These topics merit further study in order to optimize the imaging performance and facilitate further data analysis, particularly for specific radio telescopes such as the MUSER.

\section{Imaging Pipeline Design} \label{sec:imagingofmuser}
\subsection{Imaging Pipeline}
\begin{figure*}[htbp]
  \centering
  \includegraphics[width=0.9\textwidth]{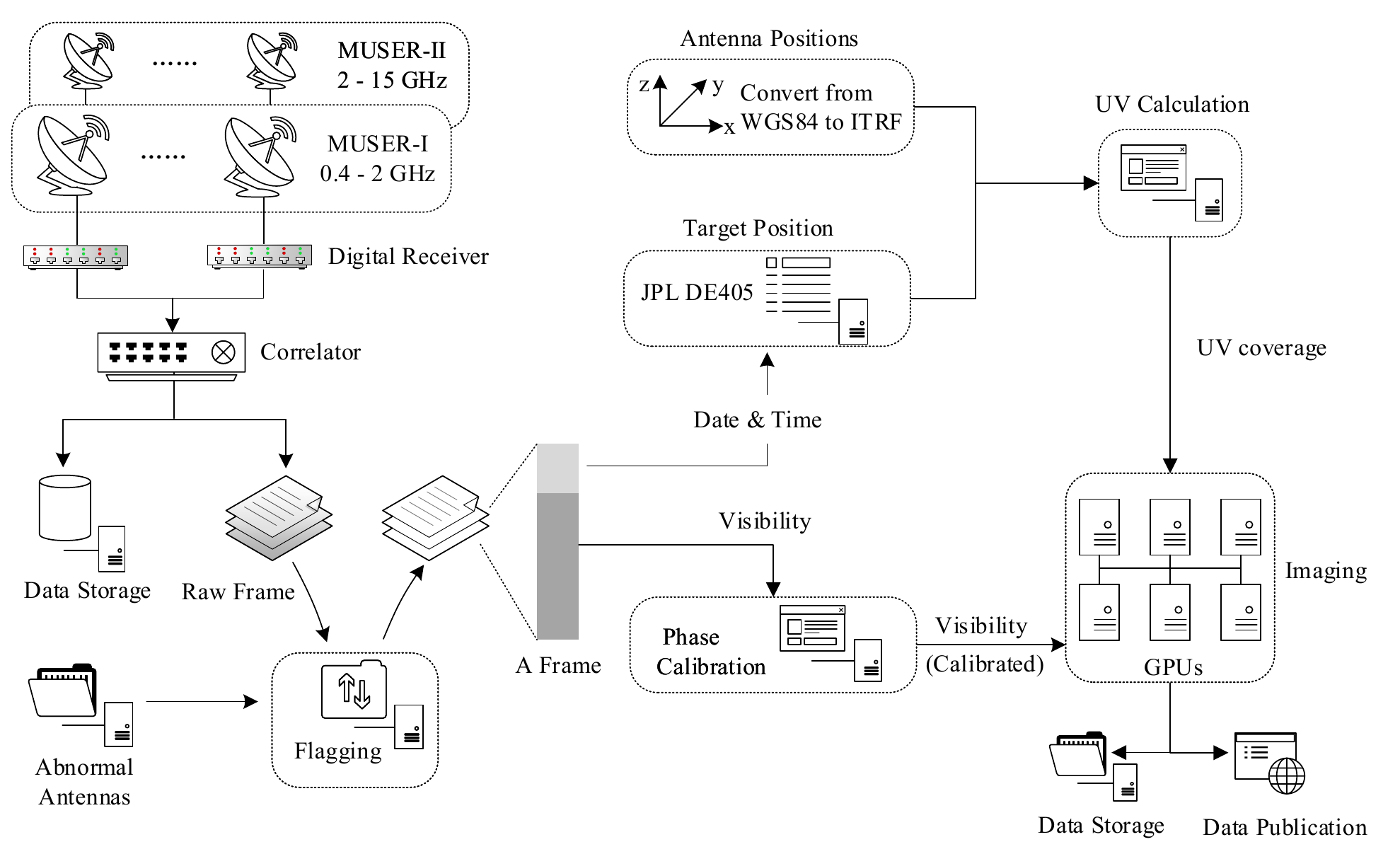}
   \centering
   \caption{MUSER imaging pipeline.}
   \label{ImagingPipeline}
\end{figure*}

The imaging of the MUSER follows the basic principles of an interferometric telescope. Currently, a data processing pipeline has been designed to automatically process the data from the MUSER. Figure~\ref{ImagingPipeline} illustrates the complete data processing procedure from the raw data acquisition, abnormal data flagging, phase calibration, UV calculation, imaging and the data storage to data publication. 

A series of data pre-processing procedures, i.e., reading the raw data, flagging, and calibration, needs to be performed before imaging. The steps of transmission delay compensation and fringe stopping are performed by the correlator directly, and phase calibration is performed by the geosynchronous satellite as per the MUSER's current operational status.

To assist the subsequent processing, a lightweight database is created for storing the information of malfunctioning antennas, weather conditions, instrument status and other observation information, which are retrieved during the data processing. An automatic flagging method based on the support vector machine (SVM) is used to flag the visibility data if there is not manual record of malfunctioning antennas \citep{hui2017auto}.
To accurately calculate the UVW and obtain high-quality dirty images, we implemented a Python package for high-precision target position computing using JPL DE405 Ephemeris, which can meet the accuracy requirements.
The imaging process that includes weighting, gridding, FFT and clean steps is executed on a GPU platform. The imaging parameters and key steps are presented in the following sections. 

\subsection{Imaging Parameters}
\subsubsection{Image size}

The image size forms an important parameter in high-performance imaging. According to the frequency and the longest baseline as per the MUSER design specifications (see Table~\ref{basicparameters}), the spatial resolution should be approximately $51.6{}''-10.3{}''$ for MUSER-I and $10.3{}''-1.4{}''$ for MUSER-II, with the corresponding fields of view (FOVs) for non-aliasing being $1.07^\circ-5.37^\circ$ and $2.15^\circ-0.29^\circ$ respectively. The final image sizes that required to be generated for different frequencies are listed in Table~\ref{Imagingsize} (three pixels are used to present a spatial resolution). In order to clean an image of a given dimension, it is necessary to have a beam pattern twice as large the image size so that a point source can be subtracted from any point in the map.

\begin{deluxetable*}{ccccc}[htbp]
\tablecaption{Imaging size specification\label{Imagingsize}}
\tablecolumns{5}
\tablenum{2}
\tablewidth{0pt}
\tablehead{
\colhead{Frequency} & \colhead{Spatial Resolution} & \colhead{Pixel Size} & \colhead{Space Size of Image} & \colhead{Image Size}\\
\colhead{(GHz)} & \colhead{(arc seconds)} & \colhead{(arc seconds)} & \colhead{(arc minutes)} & \colhead{(pixels)}
}
\startdata
$0.4-0.6$ & $51.6{}'' - 34.4{}''$ & $17.2{}''- 11.5{}''$ & $73.4{}'\times48.9{}'$ & $256\times256$\\
$0.6-1.2$ & $34.4{}'' - 17.7{}''$ & $11.5{}''-5.9{}''$ & $98.1{}'\times50.3{}$' & $512\times512$\\
$1.2-2.0$ & $17.7{}''- 10.3{}''$ & $5.9{}''- 3.4{}''$ & $100.7{}'\times58.0{}'$ & $1024\times1024$\\
$2.0-4.0$ & $10.3{}''-5.2{}''$ & $3.4{}''-1.7{}''$ & $72.5{}'\times36.3{}'$ & $1280\times1280$\\
$4.0-8.0$ & $5.2{}''-2.6{}''$ & $1.7{}''- 0.9{}''$ & $72.5{}'\times38.4{}'$ & $2560\times2560$\\
$8.0-15.0$ & $2.6{}'' - 1.4{}''$ & $0.9{}''- 0.5{}''$ & $76.8{}'\times42.7{}'$ & $5120\times5120$\\
\enddata
\end{deluxetable*}

\subsubsection{UVW Calculation} \label{subsec:uvw}

\begin{figure*}[htbp]
\gridline{\fig{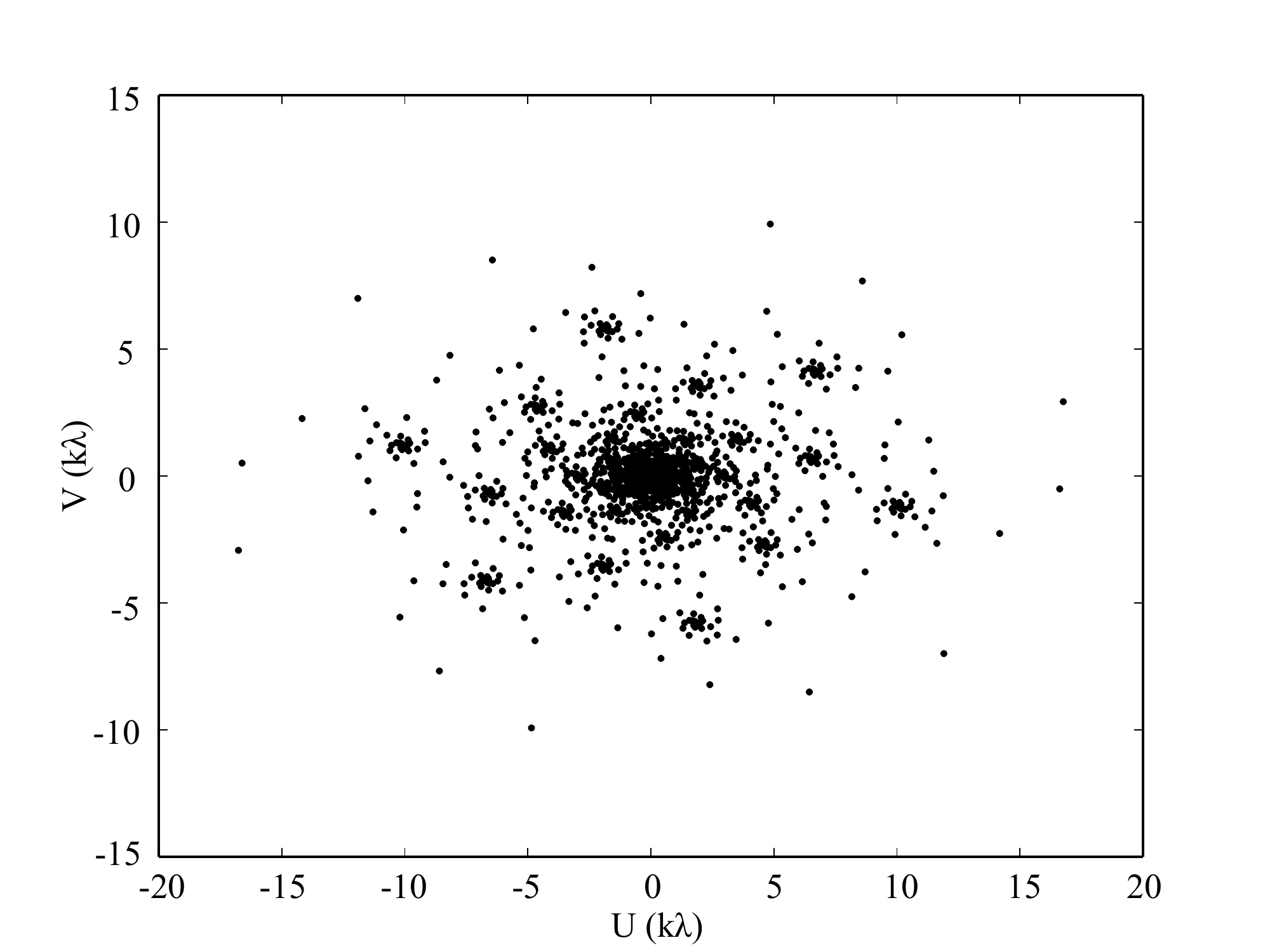}{0.4\textwidth}{(a)}
          \fig{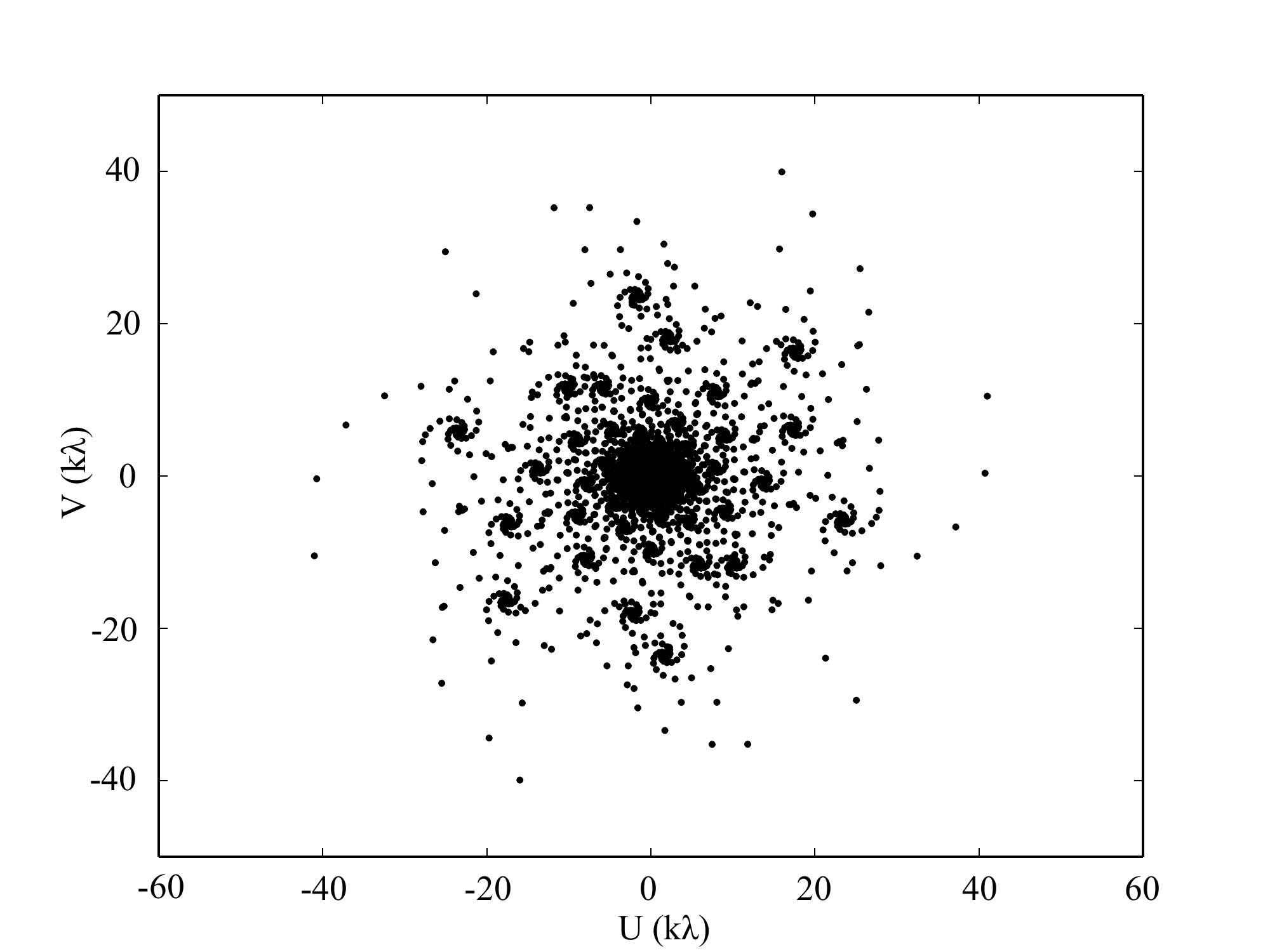}{0.4\textwidth}{(b)}
          }
\caption{UV coverage of MUSER. (a) UV coverage of MUSER-I on 2015-11-01 at 04:08:49.354161240 (UTC), frequency: 1.7125GHz, polarization: right. (b) UV coverage of MUSER-II on 2016-07-05 at 04:01:38.459449240 (UTC), frequency: 4.1875GHz, polarization: right. \label{uvdistribution}}
\end{figure*}

As all antenna positions in the MUSER are measured with the use of the Global Positioning System (GPS), which is based on the World Geodetic System WGS84, it is essential to convert this set of coordinates to International Terrestrial Reference Frames (ITRFs) before computing of the baseline vectors. The geographic coordinates of the Mingantu observational station (longitude=115.2505$^{\circ}$, latitude=42.211833333$^{\circ}$, altitude=1365.0 meters) are used during the coordinate conversion.
The variations in Earth's orientation as obtained from the IERS\footnote{\url{http://maia.usno.navy.mil}} are also considered. The UV coverage of the MUSER is shown in Figure~\ref{uvdistribution}. 

To eliminate the effects of w-term in MUSER, we chose the implementation of w-projection~\citep{muscat2014high} which can be run on GPU environment. It should be pointed out, the correction of w-term is determined by the phase error caused by ignoring the w-term, which is defined as $\left| \pi (l^2+m^2)w \right|\ll 1$, where $(l^2+m^2)$ is the synthesized field. When the phase error is small enough, we do not need to correct the w-term so as to improve the performance. A thorough analysis of the w-term is given in section 6.1.

\subsubsection{Gridding convolution function} \label{subsec:gridding}

In order to use the FFT transform to improve the imaging performance, the measured visibility must be gridded into a 2D array. For smoothing and interpolating the data, the visibility data are convolved with a suitably chosen gridding convolution function (GCF) before resampling with different weighting methods. In practice, the precise manner in which this is carried out can greatly affect the macroscopic properties of the resulting image, such as both aliasing from the sidelobes and uniform spacing of the grid \citep{Schwab1984Optimal}. Many functions, such as the pillbox, exponential, sinc and spheroidal functions, can be considered for use as the the convolution function. The spheroidal function is used in MIRIAD software \citep{sault1995retrospective}, and as per Schwab's choice of the optimal gridding method, the spheroidal function was considered as the optimal GCF. 

\begin{figure*}[htbp]
\gridline{\fig{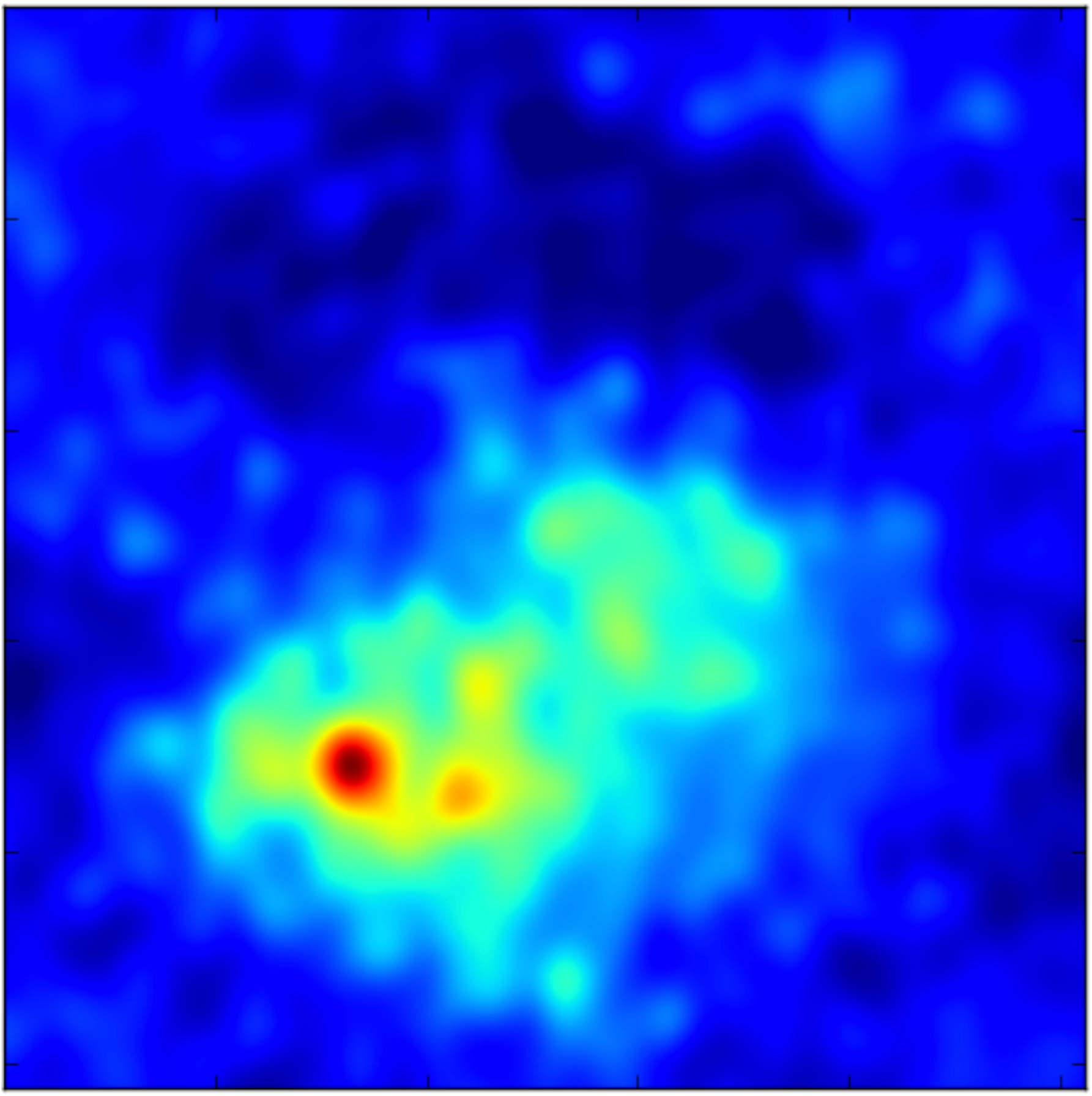}{0.3\textwidth}{(a)}
          \fig{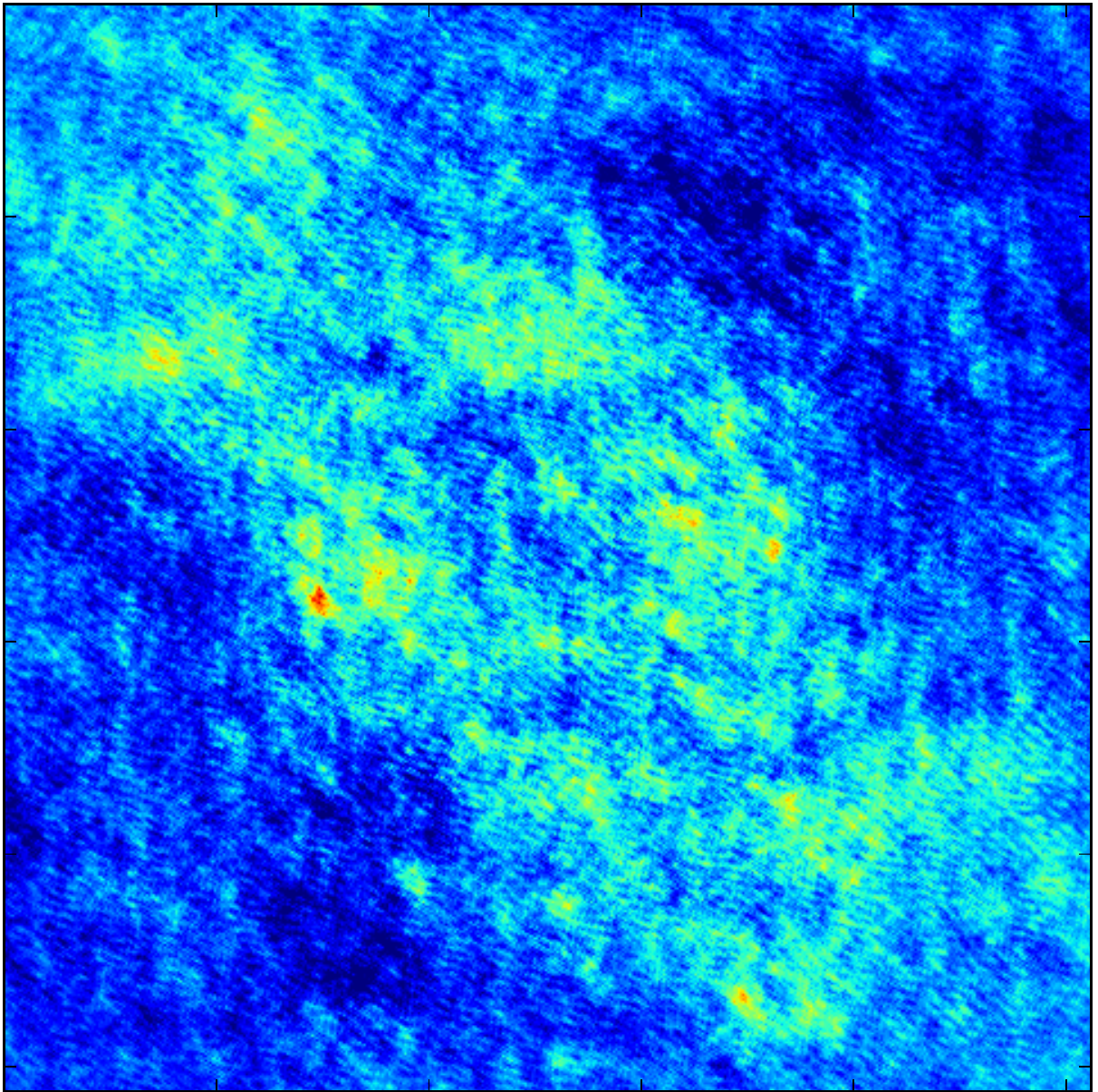}{0.3\textwidth}{(b)}
          }
\caption{(a) Dirty image from MUSER-I on 2015-11-01 at 04:08:49.354161240 (UTC) at frequency: 1.7125GHz, polarization: right. (b) 10-min-integrated dirty image from MUSER-II on 2016-07-05 at 04:04:38.459449240 (UTC) to 2016-07-05 04:15:00 (UTC) at frequency: 4.1875GHz, polarization: right. \label{dirtyimages}}
\end{figure*}

To implement the pipeline, we compared many GCFs and finally selected the truncated spheroidal function with default coefficients listed as per the study of \cite{Schwab1984Optimal} for the MUSER data gridding. That is, the visibility is convolved with a spheroidal wave functions and then gridded (uniform and natural weighting are implemented). Inverse FFT is performed after gridding for generating dirty image. The dirty images of the MUSER are shown in Figure~\ref{dirtyimages}. The dirty image from a single frame of MUSER-I is presented in Figure~\ref{dirtyimages} (a)) and Figure~\ref{dirtyimages} (b) depicts a 10-min-integrated output of MUSER-II, which can effectively improve the signal-to-noise ratio (SNR).

\section{Improved clean algorithm}

In the current MUSER data processing system, the Steer clean algorithm is used to clean the dirty image. Obviously, the application of the clean algorithm is a time-consuming iterative processes. It is necessary to estimate the probable number of iterations to improve the clean performance, rather than depending on prior experiential estimates or a given upper limit.
Moreover, as can be observed from the dirty images (Figure~\ref{dirtyimages}), there is a phase error leading to the solar disk being located off-center in the current stage, which will also affect the subsequent data processing steps such as P-angle correction. 

\begin{figure*}[htbp]
   \centering
   \includegraphics[width=\textwidth]{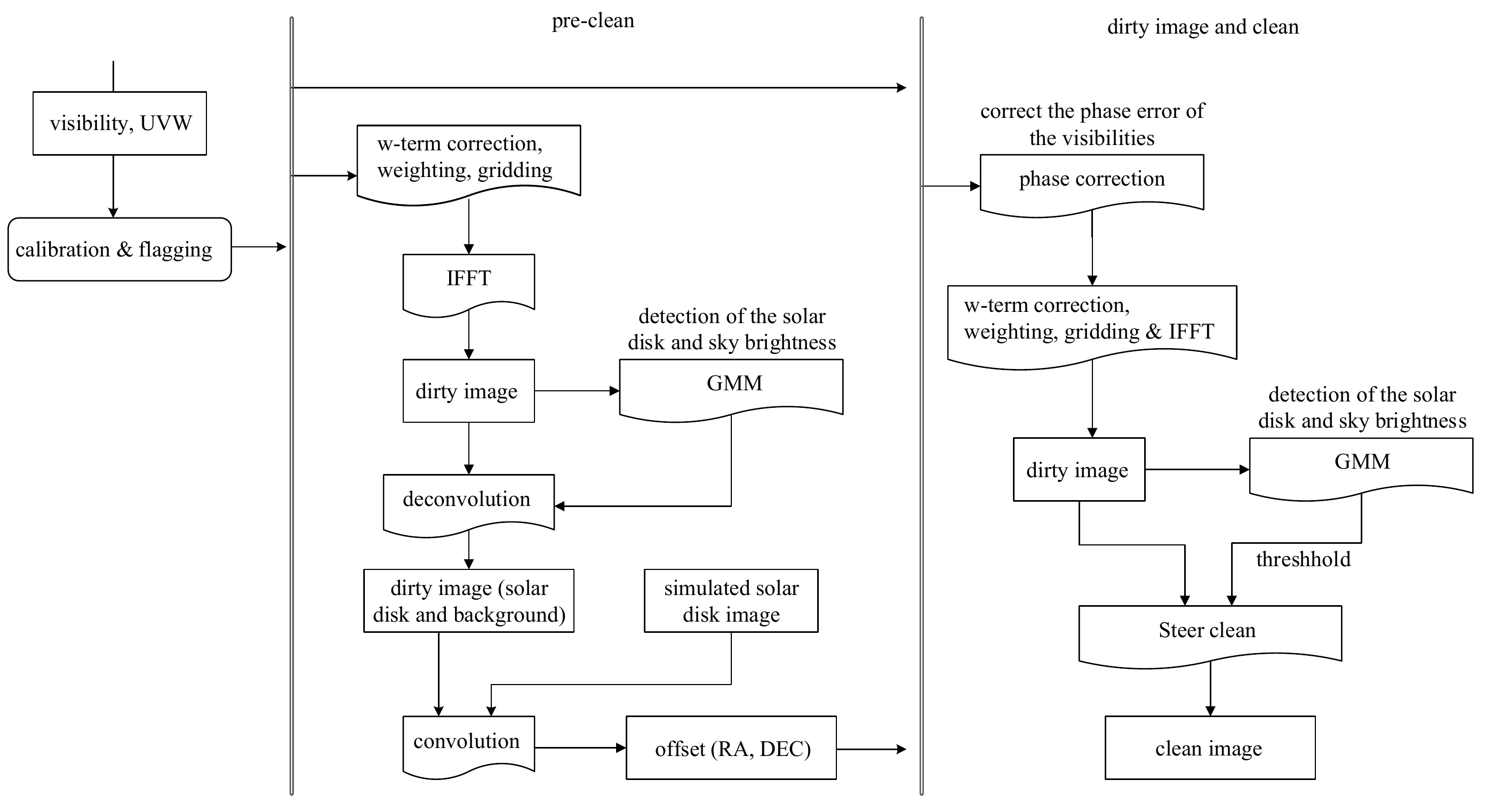}
   \centering
   \caption{Flowchart showing imaging and clean processes for the MUSER.}
   \label{FlowchartClean}
\end{figure*}

Figure~\ref{FlowchartClean} presents the improved clean process. We add a ``pre-clean'' procedure to detect the solar disk and sky brightness and then carry out the phase correction. Imaging and the improved clean are subsequently performed to obtain clean images. 

\subsection{Detection of Solar Disk and Sky Brightness}

Theoretically, there are the sky background, solar disk area and solar activity area (if there is solar activity) in the dirty image, and the brightness of these areas differs distinctly. Through analysis, the judgment of different regions reduces to a clustering problem, and data points within the same group can be modeled by a Gaussian distribution. A classical algorithm for classification and modeling, the Gaussian mixture model (GMM)  \citep{ivezic2014statistics, wright2015machine} is used in our study.

\begin{figure*}[htbp]
   \centering
   \includegraphics[width=0.9\textwidth]{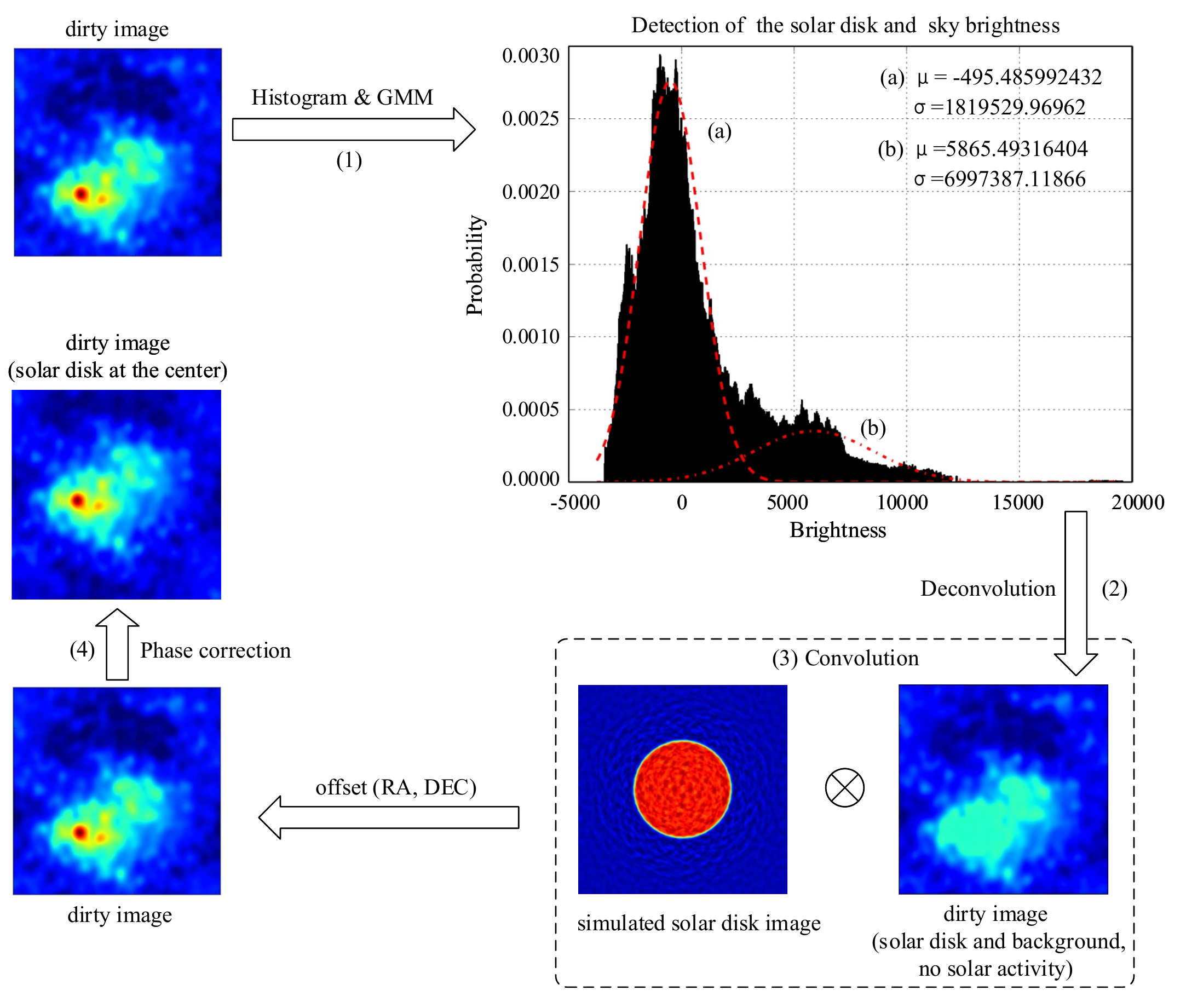}
   \centering
   \caption{Process of detecting the solar disk and sky brightness and phase correction (pre-clean), considering the dirty image of MUSER-I as an example.}
   \label{dirtyimageandGMM}
\end{figure*}

For estimating the solar disk and brightness, the data from a dirty image is divided into 2000 ranges. The histogram in Figure~\ref{dirtyimageandGMM} depicts the probability of data points in the corresponding range (considering the dirty image of MUSER-I in Figure~\ref{dirtyimages} as an example). GMM is performed to estimate the parameters of the potentially included Gaussian distributions in the histogram and the results are plotted in Figure~\ref{dirtyimageandGMM}. 
According to the brightness of different areas, $\mu$ in curve (a) denotes the maximum likelihood brightness of the sky brightness and $\mu$ in curve (b) indicates the brightness of the solar disk, and this result will be used during the clean processing.

\subsection{Disk Centering Correction}
After obtaining the estimated solar disk and sky brightness, we follow three steps for disk centering correction:
\begin{enumerate}
\item To ensure that there is no solar activity in the dirty image and to acquire more precise offset parameters, deconvolution is performed on the original dirty image until the brightness is equal or lesser than the estimated solar disk brightness (i.e., $\mu$ in curve (b)) from the GMM. This deconvolution processing ensures that the dirty image contains only the solar disk and sky background. 
\item A correlation operation is performed between the dirty image in step (1) and the simulated solar disk to obtain the center offset values (i.e., RA offset and DEC offset). 
\item The phase of the visibility data is corrected by the RA offset and DEC offset. Figure~\ref{dirtyimageandGMM} illustrates the process of this correction.

\end{enumerate}

\subsection{Clean with determined number of iterations}
\label{sect: MUSER Image Clean}
In view of the specific characteristic of solar imaging and referring to NoRH, we use a modified deconvolution algorithms, i.e., the Steer clean algorithm, to clean the dirty image acquired with the MUSER. As the number of clean subtractions and the loop gain ($\gamma$) determine the depth of the clean operation \citep{taylor1999synthesis}, which affects the image quality and imaging performance, it is crucial to optimize the loop gain and number of iterations. In this regard, the results of Very Large Array applications suggest that the reasonable compromise range of the loop gain is $0.1 \leq \gamma \leq 0.25$ \citep{taylor1999synthesis}. Through experiments, $\gamma=0.1$ is chosen as the optimal loop gain. 
For the deconvolution, the estimated sky brightness (i.e., $\mu$ in Figure~\ref{dirtyimageandGMM} curve (a): $\mu_{a}$) is used as a threshold to automatically determine the number of iterations. That is, the iteration stops when the brightness of the sky is reached. 

\section{GPU-based imaging implementation and performance}
\label{sect:Performance discussion}

\begin{figure*}[tbp]
\gridline{\fig{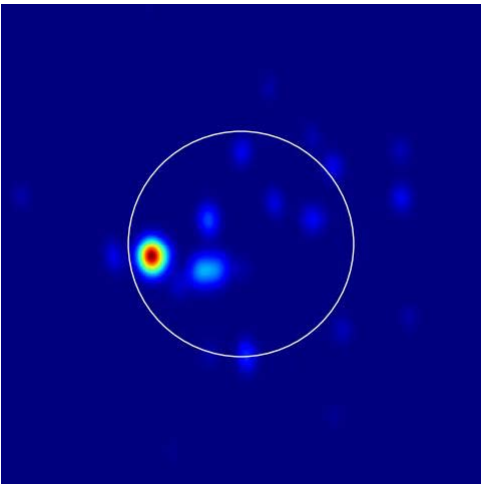}{0.245\textwidth}{(a)}
          \fig{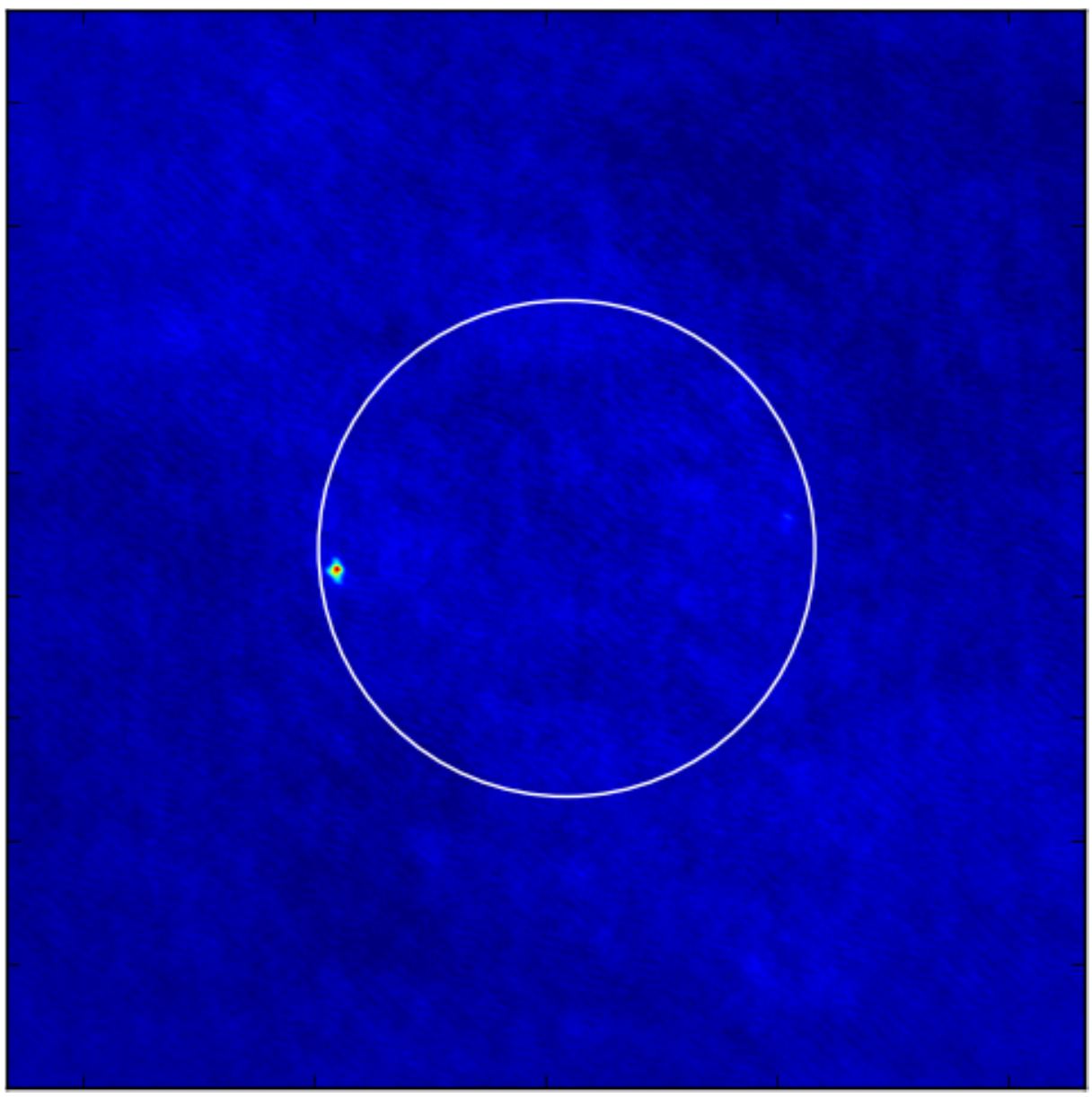}{0.245\textwidth}{(b)}
          \fig{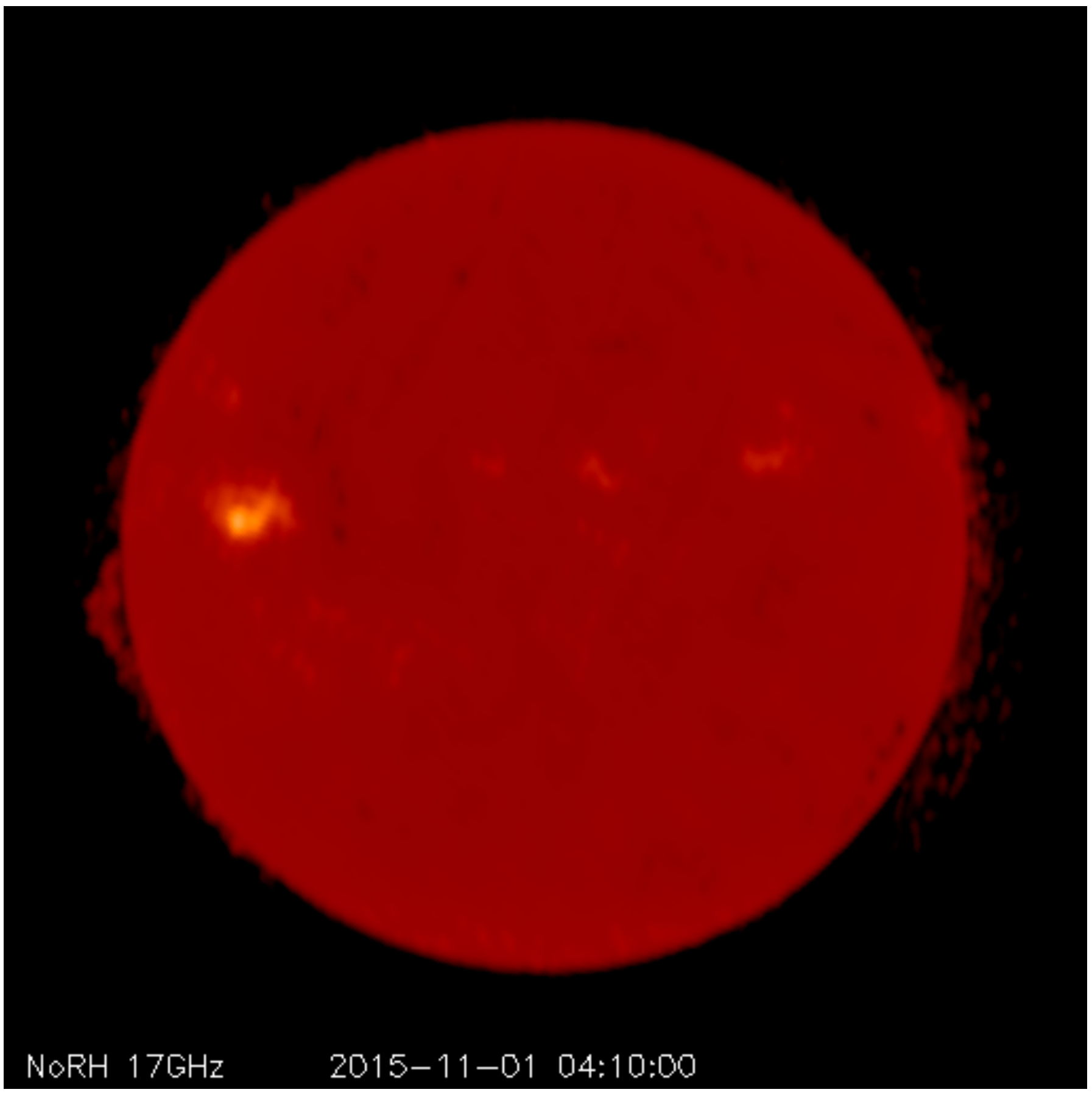}{0.245\textwidth}{(c)}
          \fig{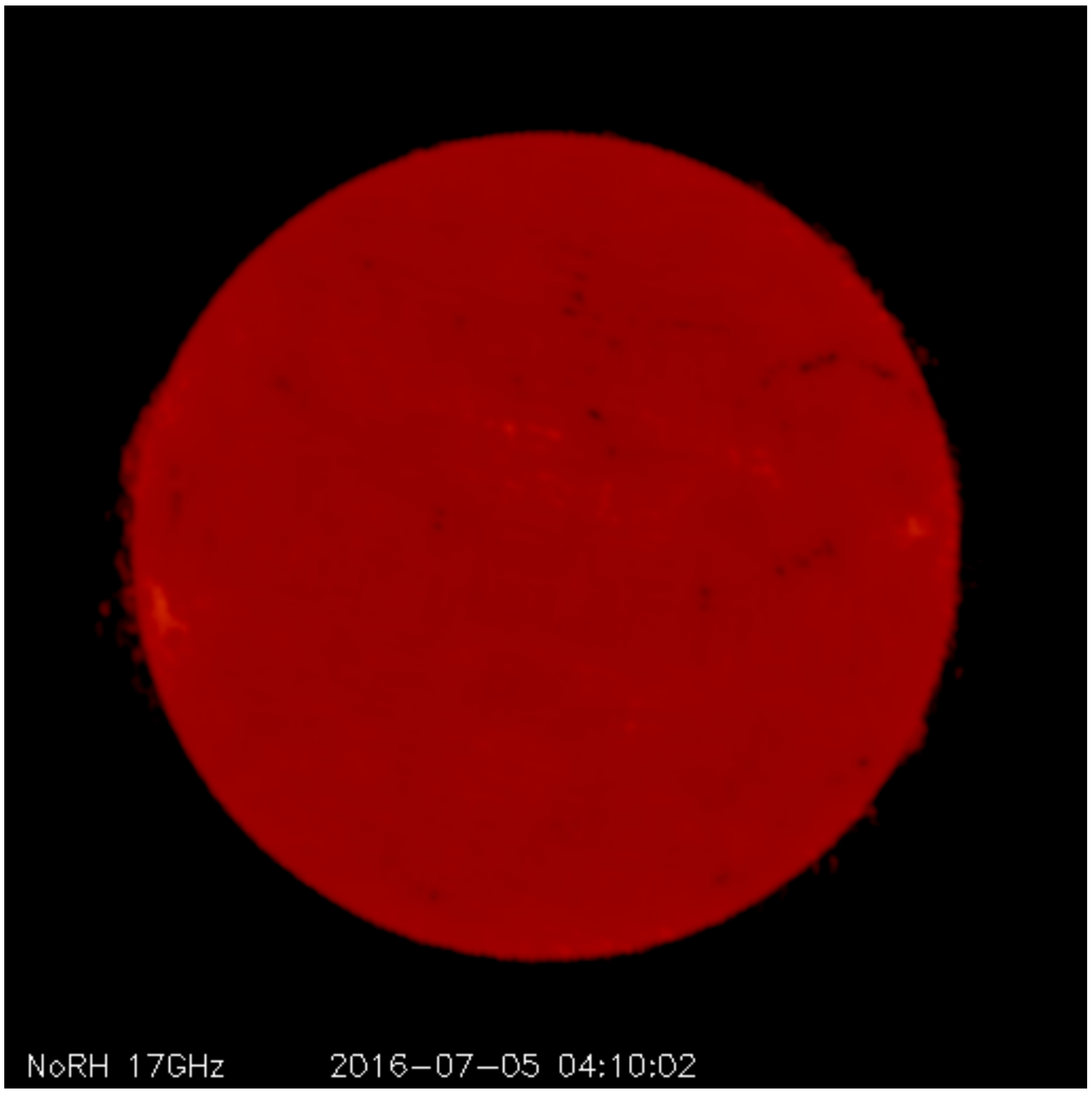}{0.245\textwidth}{(d)}
          }
\caption{(a) Clean image from MUSER-I on 2015-11-01 at 04:08:49.354161240 (UTC), frequency: 1.7125 GHz, polarization: right. (b) 10-mins-integrated clean image of MUSER-II from 2016-07-05 04:04:38.459449240 (UTC) to 2016-07-05 04:15:00 (UTC), frequency: 4.1875GHz, polarization: right. (c) Clean image from Nobeyama on 2015-11-01 at 04:10:00 (UTC), frequency: 17 GHz (R+L). (d) Clean image from Nobeyama on 2016-07-05 at 04:10:02 (UTC), frequency: 17 GHz (R+L)
\label{cleanimages}}
\end{figure*}

CUDA is a GPU programming model developed by NVIDIA, and it provides a programming interface to utilize the highly-parallel nature of GPUs and mask the complexity of controlling GPUs. For the MUSER data processing, the data preparation is carried out in the CPU environment, and here, we focus on the computation-intensive processes such as gridding, FFT and clean, which are accelerated with the use of GPUs. Algorithm~\ref{Algorithm1} below describes the improved clean process. The clean images are shown in Figure~\ref{cleanimages} together with the images from NoRH as a comparison.

All the codes used in our study is available at \url{https://github.com/astroitlab/museros}. Python is used as the programming language as it is prevalent in astronomical data processing. More importantly, it's really convenient to use packages such as PyCUDA and Scikit-CUDA, which provide the access to the CUDA parallel computation application programming interface (API). The performances achieved for various key steps are listed in Table~\ref{performance}, considering two representative image sizes as an example. The elapsed time of Steer clean is indicated for one iteration, and the number of iterations for cleaning the dirty images in Figure~\ref{dirtyimages} (a) is 39 and 33 for  Figure~\ref{dirtyimages} (b). That is, the total time for generating a dirty image is about 0.26 seconds for MUSER-I and 0.8 seconds for MUSER-II, and for Steer clean, a further about 1.138 s is needed for MUSER-I and 1.45 s is needed for MUSER-II. 

\section{Discussion}
The imaging pipeline presented in this study meets the requirements of automatically processing the massive observational data from the MUSER in the current stage. However, there are still some issues that need to be further discussed. 

\subsection{w-term}
W-term is an inevitable problem in the MUSER data processing. Actually, the MUSER has the specifications of small antenna apertures, long baselines and long wavelengths. Therefore, the non-coplanar baselines effect has to be considered so as to obtain high quality observational results. 

According to the previous studies~\citep{Thompson2008}, w-term may be ignored and a two-dimensional Fourier transform can be used when the term $2\pi w(\sqrt{1-l^2-m^2}-1)$ is much less than unity. If the antennas track the source under observation down to low elevation angles, the values of $w$ can approach the maximum spacings that is approximately the maximum length of baseline. The the maximum value of the term is expressed as $\frac{B\lambda}{D^2}$, where B is the length of baseline, $\lambda$ is the wavelength and $D$ is the antenna diameter. For the observation of the MUSER with 450 MHz and the maximum baseline length of 3000 meters, the value of w-term would be 140.625, which is much larger than unity. Obviously, w-term needs to be more explicitly dealt with or the operation of the instrument will be greatly restricted.

\begin{algorithm}[H]
    \begin{algorithmic}[1]
    \caption{MUSER Clean} 
    \label{Algorithm1}
        \Require Visibilities and UVW (Phase corrected, Flagged)
        \Require weighting mode: $weight\_mode$, natural or uniform
        \Ensure clean images
        \Function {void $wgtGrid\_kernel$}{float2 *Grd, int *cnt, int nu, int mode}
            \State int iu = blockDim.x*blockIdx.x + threadIdx.x;
            \State int iv = blockDim.y*blockIdx.y + threadIdx.y;
            \State int u0 = 0.5*nu; \Comment{Applying the weight to half the grid (as the data is Hermitian)}
            \If {$(iu >= u0 \&\& iu < nu \&\& iv < nu)$}
                \If {$(cnt[iv*nu+iu]!= 0)$}
                    \State int ind = iv*nu+iu;
                    \State float foo = cnt[ind];
                    \State float wgt;
                        \If {(mode == 1)}
                            \State wgt = 1./foo;  \Comment{Uniform Weighting}
                        \Else 
                            \State wgt = 1; \Comment{Natural Weighting}
                        \EndIf
                    \State Grd[ind].x = Grd[ind].x*wgt;
                    \State Grd[ind].y = Grd[ind].y*wgt;
                \EndIf
            \EndIf
        \EndFunction
        \State
        \Function {dirty\_image}{$visibilities, UVW, weight\_mode$}
        	\State W-term correction (w-projection);
            \State Calculating the weight and gridding: $wgtGrid\_kernel$;
            \State Invert Fast Fourier Transform to get the dirty image, with the sun disk at the center;
            \State \Return {dirty image, dirty beam and clean beam: $gpu\_dirty$, $gpu\_dpsf$, $gpu\_cpsf$};
        \EndFunction
        \State
        \Function {pre\_clean}{$disk\_model, dirty\_image$} \Comment{Obtaining the phase errors}
            \State Apply GMM to a dirty image ($dirty\_image$) and get the estimated solar disk and sky brightness;
            \State Picking out points with brightness-values greater than the estimated solar disk in the dirty image: h\_dirty;
            \State Calculating the offset parameters using the the disk model and the h\_dirty: $sun\_disk\_offset(disk\_model, h\_dirty)$;
            \State \Return {ra\_offset, dec\_offset};
        \EndFunction
        \State
        \Function {improved\_clean}{$dirty\_image, band, channel$}
            \State Obtaining the phase error: (ra\_offset, dec\_offset) = $PRE\_CLEAN()$;
            \State Phase correction: correct the phase error of the visibility data;
            \State Generating the dirty image: $DIRTY\_IMAGE()$;
            \State Find the strength and position of the peak in the dirty map: ($imax$, $gpu\_max\_id$);
            \While{$imax>\mu_{a}$}   \Comment{loop until: $imax$ (the maximum brightness) $\leq\mu_{a}$ (the sky brightness)}
                \State Steer clean;
                \State Find the strength and position of the peak in the dirty map: ($imax$, $gpu\_max\_id$);
            \EndWhile
            \State \Return {clean images};
        \EndFunction
	\end{algorithmic}
\end{algorithm}

\begin{deluxetable*}{cccc}[h]
\tablecaption{Elapsed time of key steps in MUSER imaging \label{performance}}
\tablecolumns{4}
\tablenum{3}
\tablewidth{0pt}
\tablehead{
\colhead{Subtask No.} &
\colhead{Subtask} &
\colhead{Frequency/Pixel} &
\colhead{Elapsed time (s)}
}
\startdata
  1 &     Data preparation        &  0.4 GHz - 2.0 GHz & 0.06161 \\
    & (Calibration and flagging)  &  2.0 GHz - 15 GHz & 0.13885\\
  \hline                          
  2 & Gridding, weighting and FFT & 1024$\times$1024 & 0.20664\\
    &                             & 2560$\times$2560 & 0.66279\\
  \hline
  3 & Detection of solar disk and sky brightness   & 1024$\times$1024 & 0.44671\\
    &  (GMM)                 & 2560$\times$2560 & 0.46264\\
  \hline
  4 & Phase correction       & 1024$\times$1024 &  0.01773 \\
    &                        & 2560$\times$2560 &  0.02987 \\
  \hline
  5 &      Pre-clean         & 1024$\times$1024 &  0.67108 \\
    &   (Subtask 2+3+4)      & 2560$\times$2560 &  1.15530 \\
  \hline
  6 &     Steer clean          & 1024$\times$1024 &  0.01773 \\
    &  (One iteration)         & 2560$\times$2560 &  0.02987 \\
  \hline
  7 &  Generating dirty image  & 1024$\times$1024 &  0.26825 \\
    &  (Subtask 1+2)           & 2560$\times$2560 &  0.80164 \\
  \hline
  8 &  Generating clean image  & 1024$\times$1024 &  3.21551 \\
    &  (Subtask 1+2+3+5+6*NOI)     & 2560$\times$2560 &  4.85529 \\
  \hline
\enddata
\tablecomments{NOI: Number of iterations. Testing environment: Intel(R) Xeon(R) CPU, 6 cores, 2.10GHz. GPU: NVIDIA Corporation GM200.}
\end{deluxetable*}

In general, the w-term can be ignored when
$\left| \pi (l^2+m^2)w \right|\ll  1$.
We can possibly allow it to be as high as 0.1 radian without introducing serious errors in the image~\citep{taylor1999synthesis}. The observational target of the MUSER is the Sun and the radius of the Sun is about $32''$, that is, $(l^2+m^2)$ can be roughly $256''$. To guarantee the valid of the equation, the value of the $w$ must be less than 1504.3 (unit: number of wavelength). Obviously, for the observation with different frequency and different target position, we have to calculate the value of the w-term and finally determine the correction of the w term. If the value of the $w$ is less than 1504.3, the correction is not necessary. 

We finally realize that the corrections for the w-term is necessary for the MUSER in most cases (about more than 300 days in a year). We calculated the values of w-term at 12:00 am (China Beijing time) every day in the year of 2017 respectively. The maximum value of the w-term (about 2821.55, unit: number of wavelength) in a year appeared on Dec 11 instead of the winter solstice. The minimum value (only 0.000357) of the w-term is on Jun 30. Even on the best day (Jun 30) in a year, only the data in four hours (10:10 am - 14:10 pm) can be directly processed without w-term correction. 

\subsection{Calibration}. 

Calibration is one of the most important issues in interferometric imaging. Currently, two new 20-m diameter antennas are under construction and they are to be used for calibration in the near future. Prior to application of these antennas, we have no effective approach to calibrate the phase and amplitude errors accurately. To ensure that the MUSER generates scientific results as early as possible, a satellite is used to calibrate phase error in the current stage. 

Simulations in our previous work proved that the observed image appears a translation when the satellite is subject to position errors. That's to say, the phase error that leads to the deviation of the solar disk is caused by the pointing error of the satellite. Therefore, it is necessary to further correct the phase error in the pre-clean processing step in the current stage. Current data processing procedure has no absolute calibration with respect to amplitude, scientists can use these data to carry out their scientific research in other specific fields as well.

\subsection{Performance and Deployment}
From Table~\ref{performance}, we note that the current data processing affords satisfactory performance with the use of GPU technology. we note that the elapsed time of generating a dirty image of 1024$\times$1024 is about 0.268 seconds and 0.8 seconds for a 2560$\times$2560 dirty image. 

Full image processing involving dirty image generation, pre-clean and Steer clean would take about 3.22 seconds for a 1024$\times$1024 image and 4.86 seconds for a 2560$\times$2560 image with the use of one machine with one GPU card. Significantly, the comparison experiments of generating clean image using CASA by processing UVFITS file of the MUSER show that the performance of Steer clean based on GPU technology is about 2--8 times faster than the CASA software package, and this performance advantage becomes more obvious especially in processing the data of MUSER-II.

The data processing pipeline has been deployed both for routine observational monitoring and scientific data processing of the MUSER. For real-time monitoring, the aforementioned distributed data processing framework -- Opencluster is used to schedule the pipeline on a high-performance CPU/GPU cluster. The current monitoring system could produce 16 cross-correlograms, 16 spectrum diagrams and 16 dirty images of one band in every 5 seconds for MUSER-I and MUSER-II respectively. The performance is enough for the requirements of real-time monitoring. 

For the requirement of the scientific data processing, not all observational data would be processed. The pipeline has also been integrated into a command line system, which is convenient for post data processing as it can operate separately on a single machine. In common, the scientists would process the data at a specific time according to their research goal. According to Table~\ref{performance}, the current data processing system can already generate high-quality images with a high-performance.

\subsection{Limitations}

Detecting the solar disk and sky brightness is indispensable for automatically determining the number of iterations and optimizing the clean performance. However, the current approach strongly depends on the signal-to-noise ratio (SNR) of the image. 

According to the results of data processing, the proposed approach of detecting the solar disk and sky brightness is well-suited for high-SNR images because the peak position in the histogram could be correctly recognized. Therefore, this approach is to a certain extent limited for processing snapshot observational data (3 ms/frame) especially in the data processing of MUSER-II.

In addition, a high-performance CPU/GPU cluster is indispensable to process the massive data generated by the MUSER. The current performance level can meet the requirements of a monitoring system which refresh the observational results in every 5 seconds. However, the power consumption needs to be considered because the power supply for a distant observational station is limited. 

\section{Conclusions and Future work}
\label{sect:conclusion}
We optimized and implemented a high-performance data processing pipeline for the MUSER in this study. 
The preliminary imaging results prove the feasibility of the proposed processing approach. We specifically draw attention to the phase correction in current status and the detection of the solar disk and sky brightness based on the Gaussian mixture model, which automatically determines the number of iterations for clean algorithms. More importantly, the performance of Steer clean based on GPU is far better than the commonly used CASA software.

We have already begun studies on new imaging algorithms such as compressed sensing and so on. Meanwhile, to further improve the imaging quality, it is necessary to further study effective flagging methods, visibility integration, calibration and clean algorithms. 

Actually, as same as other new telescope systems, the construction of the MUSER is a long-term effort. The proposed approaches accelerated the scientific data output of the MUSER. 
Regardless of limitations of the current MUSER operation, we believe that the scientific output of the MUSER is promising and that the MUSER can significantly contribute to astronomical observations in the near future.

\begin{acknowledgements}
This work is supported by the National Key Research and Development Program of China (2016YFE0100300), the Joint Research Fund in Astronomy (No. U1531132, U1631129, U1231205) under cooperative agreement between the National Natural Science Foundation of China (NSFC) and the Chinese Academy of Sciences (CAS), the National Natural Science Foundation of China (No. 11403009 and 11463003).

We appreciate the suggestions from Prof. Monique Pick of Paris Observatory and Prof. Hiroshi Nakajima of the National Astronomical Observatory of Japan. The authors also gratefully acknowledge the helpful comments and suggestions of the reviewers.
\end{acknowledgements}

\bibliography{muser} 
\bibliographystyle{aasjournal}

\end{document}